\documentclass[nohyper,11pt,letterpaper]{JHEP3}
\usepackage{graphicx}
\usepackage{amsmath}
\usepackage{amssymb}
\usepackage{amsthm}
\usepackage{amsfonts}

\newcommand{\be}{\begin{equation}}
\newcommand{\eb}{\end{equation}}

\newcommand{\ba}{\begin{eqnarray}}
\newcommand{\ab}{\end{eqnarray}}

\newcommand{\stn}{\frac{\text{S}}{\text{N}}}

\title{Power Spectrum and Non-Gaussianities in Anisotropic Inflation}

\author{Anindya Dey\\
Theory Group, Department of Physics and Texas Cosmology Center\\ The University of Texas at Austin,
TX 78712.
\\ E-mail: \email{anindya@physics.utexas.edu}}

\author{Ely D. Kovetz \\
Theory Group, Department of Physics and Texas Cosmology Center\\ The University of Texas at Austin,
TX 78712.
\\ E-mail: \email{elykovetz@gmail.com}}
\author{Sonia Paban\\
Theory Group, Department of Physics and Texas Cosmology Center\\ The University of Texas at Austin,
TX 78712.
\\ E-mail: \email{paban@physics.utexas.edu}}

\abstract{We study the planar regime of curvature perturbations for  single field inflationary models  in an axially symmetric Bianchi I background. In a theory with standard scalar field action, the power spectrum for such modes has a pole as the planarity parameter goes to zero. We show that constraints from back reaction lead to a strong lower bound on the planarity parameter for high-momentum planar modes and use this bound to calculate the signal-to-noise ratio of the anisotropic power spectrum in the CMB, which in turn places an upper bound on the Hubble scale during inflation allowed in our model. We find that non-Gaussianities for these planar modes are enhanced for the flattened triangle and the squeezed triangle configurations, but show that the estimated values of the $f_{\rm NL}$ parameters remain well below the experimental bounds from the CMB for generic planar modes (other, more promising signatures are also discussed). For a standard action, $f_{\rm NL}$ from the squeezed configuration turns out to be larger compared to that from the flattened triangle configuration in the planar regime. However, in a theory with higher derivative operators, non-Gaussianities from the flattened triangle can become larger than the squeezed configuration in a certain limit of the planarity parameter.}

\keywords{Anisotropy, Power Spectrum, Bi-spectrum}

\received{???????? ?st, 2013} \accepted{???????? ?th, 9999}
\preprint{\hepth{yymmnnn}\\UTTG-35-13 \\TCC-029-13 }

\begin{document}
\section{Introduction}
  In view of the first cosmological data release by Planck \cite{PlanckNG}, it appears that all observations remain consistent with the general predictions of a standard single field inflationary scenario. However, it remains an important exercise to analyze scenarios which depart from the standard single-field model in dynamically non-trivial ways and that are still consistent with the Planck data, while providing predictions that could be tested with future data. One such scenario consists of having a single scalar inflaton evolve in an axially symmetric Bianchi I (Kasner-de Sitter) space-time \cite{Dey:2011mj, Kim:2011pt}. The standard single field scenario takes the initial metric to be a FRW isotropic metric, and the so-called magic of inflation is supposed to be that the dependence of the observables on these initial conditions disappears after few e-foldings. This paper continues the work of \cite{Dey:2011mj, Kim:2011pt} in further testing this conjecture. It was shown in  \cite{Dey:2011mj, Kim:2011pt} that the scalar perturbations in such a model admit a WKB solution for modes in a certain high-momentum regime. A crucial point of this computation was to show that matching this WKB solution with the de Sitter solution leads to a description of the late time dynamics of the curvature perturbation in terms of an excited state on the Bunch-Davies vacuum. Although there exists a substantial literature on the enhancement of the local bispectrum from excited states \cite{Holman:2007na, Ganc:2011dy, Agullo:2010ws, Chen:2006nt,  Chen:2008wn, Meerburg:2009ys, Kundu:2011sg,Chialva:2011hc, Agarwal:2012mq, Kundu:2013gha}, the excited state in our case does not lead to any large enhancement of non-Gaussianity for a generic mode in the regime where the WKB solution is valid. This agrees with the results of \cite{Flauger:2013hra, Aravind:2013lra} which take into account back reaction effects of such excited states. In our case, such a result is expected since the WKB approximation is only valid for high-momentum modes. Aside from a certain corner of the momentum space which will play a prominent role in this work, the state for such modes is very close to the Bunch-Davies vacuum and any deviation in the power spectrum or non-Gaussianity is exponentially suppressed. For the case of the non-planar regime of momenta ($k_x \sim k_y \sim k_z$), this was explicitly shown in \cite{Dey:2011mj}. \\
     \indent In \cite{Dey:2012qp}, a particularly interesting regime of the high-momentum modes with $k_x \ll k_y,k_z$ -- called ``planar modes" -- was studied for a single-field inflationary scenario with a standard action. It was shown that physical observables at late times for these planar modes depend strongly on the ``planarity parameter" $s=\frac{k_x}{k}$. For example, the power spectrum for the scalar perturbation has a pole at $s=0$, which signals the breakdown of perturbation theory as one approaches the limit $s \to 0$. It was also shown in  \cite{Dey:2012qp} that the power spectrum approached the standard form in the regime $\frac{sk}{H} \sim O(1)$ and this was the regime of $s$ in which the physical observables were computed. \\
     \indent However, it is interesting to consider the physics of planar modes at lower values of $s$.  In this work, we perform a back reaction computation and obtain a theoretical lower bound on $s$.  This is then used to compute the power spectrum and bi-spectrum of the scalar perturbation for relevant configurations -- firstly, for a single field model with a standard action and then a single field model with a dimension 8 operator. \\
     \indent The paper is organized in the following fashion. Section \ref{mot} provides the necessary background  for the model considered in this paper. Section \ref{br} details the back reaction computation that gives a lower bound on the planarity parameter $s$. Section \ref{stanact} and section \ref{hdact} respectively describe the power spectrum and the bi-spectrum computation for the case of standard action and an action with higher derivative  operator, with some details deferred to the appendix. Finally, Section \ref{mainres} summarizes the important analytical formulae and the numerical results and includes a discussion on the detectability of our model.

\section{The Model}\label{mot}
A Bianchi I geometry of the Kasner-de Sitter type appears naturally in a theory of Einstein gravity with a minimally coupled single scalar field
\begin{equation}
S=\frac{M^2_p}{2}\int d^4x \sqrt{-g} R + \int d^4x \sqrt{-g}\left(-\frac{1}{2}g^{\mu\nu}\partial_{\mu}\phi\partial_{\nu}\phi -V(\phi)\right)
\end{equation}
 The metric in the above action (which we shall refer to as the background metric) is chosen to be an axially symmetric version of the
Bianchi I metric
\begin{equation}
ds^2=-dt^2 + e^{2\rho}(dx)^2 +e^{2\beta}(dy^2+dz^2) \label{metric}
\end{equation}
\indent The evolution of such a scalar field in this anisotropic geometry has been discussed in much detail in earlier works \cite{Dey:2011mj,Kim:2011pt, Dey:2012qp,Gumrukcuoglu:2007bx, Gumrukcuoglu:2008gi}.

In contrast to the FRW case where one has a single Hubble constant, we have two Hubble constants, which we choose to define as follows
\begin{equation}
H=\frac{\dot{\rho}+2\dot{\beta}}{3}, \hspace{2ex} h=\frac{\dot{\rho}-\dot{\beta}}{\sqrt{3}}
\end{equation}
The classical dynamics of the system specified by the action (2.1) involves a strongly anisotropic expansion at early times (parametrized by $h$) followed by eventual isotropization at a time-scale $t \approx t_{iso}=\frac{M_p}{\sqrt{V}}$. Ensuingly, for $t \gg t_{iso}$, the universe enters a phase of de Sitter expansion.\\

The second Hubble constant $h$  is a measure of the rate of anisotropic expansion. Note that it vanishes in the isotropic limit ($\dot{\rho}=\dot{\beta}$) so that we are left with a single Hubble constant.\\

The independent Einstein's equation and the equation of motion for the scalar field can be recast into the following set of equations using $H$, $h$ and $\phi$ as variables.
\begin{eqnarray}
\dot{H}+3H^2 &=&V(\phi)/M^2_p\\
3H^2-h^2 &=& \frac{1}{M^2_p}(\frac{1}{2}\dot{\phi}^2 + V(\phi))\\
\ddot{\phi}+3H\dot{\phi}& = & -V'(\phi)   
\end{eqnarray}

Differentiating the second equation w.r.t. time and using all three equations, one obtains the following time-evolution equation for $h$
\begin{equation}
h(\dot{h}+3Hh)=0
\end{equation}
For the solution to have an anisotropic phase of expansion we need $h \neq 0$, which implies that
\begin{equation}
\dot{h}+3Hh=0.
\end{equation}

For a general $V(\phi)$, one can only obtain approximate solutions to the above system of equations. However, in the special case of a pure cosmological constant, $V$, the coupled differential equations for $H,h$ and $\dot{\phi}$ can be exactly solved
\begin{eqnarray}
H& =& \sqrt{\frac{V}{3 M^2_p}} \, \coth(\frac{\sqrt{3V}t}{M_p}) = H_I \,  \coth(\frac{{\sqrt{3V}t}}{M_p}) \nonumber\\
h & = & \pm \sqrt{\frac{V}{M^2_p}} \, \frac{1}{\sinh({\sqrt{3V}t/M_p})}  \label{sip}\\
\dot{\phi} &= & 0  \nonumber
\end{eqnarray}
where we have defined $H_I=\sqrt{\frac{V}{3 M^2_p}}$. The explicit solution for the scale factors are given as
\begin{eqnarray}
\rho& =& \frac{1}{3}\ln{\tanh^{2}{\frac{3H_I t}{2}}\sinh{3H_I t}} \nonumber \\
\beta& =& \frac{1}{3}\ln{\frac{\sinh{3H_I t}}{\tanh{\frac{3H_I t}{2}}}} \label{classeom}
\end{eqnarray}

In the above solution, the constants have been chosen such that the metric approaches a Kasner solution in the limit $t \rightarrow 0^+$. The $\pm$ sign in the solution of $h$ indicates two different branches in the solution space (distinguished, among other things, by their behavior in the Kasner limit). It turns out that only for the positive branch, one can impose initial conditions on the cosmological perturbations at early times via the usual WKB approximation \cite{Gumrukcuoglu:2008gi}. This is the class of backgrounds we shall focus on for the rest of the paper.\\

In the Kasner limit ($t \to 0^+$), the metric reduces to the following form,
\begin{equation}
ds^2_{Kasner}=-dt^2 + (\frac{\sqrt{V}t}{M_p})^2 (dx)^2 + (dy^2 +dz^2)
\end{equation}
with $\dot{\rho}=\frac{1}{t}$, $\dot{\beta}=0$.\\

The cosmological perturbations evolve at early times in this gravitational background. The solutions for the background equations of motion suggest that the universe starts its evolution with a very strong anisotropy ($h\rightarrow \frac{1}{t}$ at early times) which is smoothed out very quickly by the inflaton potential. The universe then enters a phase of usual isotropic inflation. We denote the scale factor during the isotropic phase ($t \gg t_{iso}$), corresponding to the Hubble constant $H_I$, as $a(t)$ (note that in order to simplify the notation we omit the subscript $I$ whenever it is evident that we are discussing the late time behavior). \\

  Given a non-trivial $V(\phi)$, the slow-roll condition ($\ddot{\phi} \approx 0$) forces  $V(\phi)$ to be nearly constant at early times, as long as $H\dot{\phi}^2 \rightarrow 0$. All common inflaton potentials obey this condition and therefore the above solution (\ref{sip}) is valid even for a non-constant potential in the $t \rightarrow 0^+$ limit. 

 Consider, for example, the case of massive chaotic inflation \cite{Gumrukcuoglu:2007bx} with $V(\phi)=\frac{m^2\phi^2}{2}$.  In this case $H,h$ and $\phi$ have the following asymptotic forms at early times
\begin{eqnarray}
H & = & \frac{1}{3t}\left[1+\frac{m^2\phi_0^2t^2}{2M^2_p}+O(m^4t^4)\right]\\
h&=&\frac{1}{3t}\left[1-\frac{m^2\phi_0^2t^2}{4M^2_p}+O(m^4t^4)\right] \\
\phi & = & \phi_0 \left[1-\frac{m^2t^2}{4}+O(m^4t^4) \right] 
\end{eqnarray}
In this case, we have $H\dot{\phi}^2 \approx t \rightarrow 0$, so that $V$ is essentially constant at early times.\\

\section{Bound on Planarity Parameter from Back Reaction}\label{br}
As mentioned in the Introduction, the late-time dynamics of the curvature perturbation in this model of inflation may be inferred by matching the early time WKB solution for scalar perturbations with late time de Sitter solution at some intermediate time $t=t_{*}$. The relevant computation has been worked out in detail in earlier papers  \cite{Kim:2011pt, Dey:2012qp} for wave-numbers in the planar as well as the non-planar regime. Here we simply quote the result for the former which is the subject of our study. 

For  scalar perturbations in the planar regime, we find that the de-Sitter dynamics is characterized by the following Bogoliubov-transformed Bunch Davies state
\begin{equation}
\begin{split}
&\phi = A_{+} \phi_{BD} + A_{-} \phi^{\star}_{BD}\\
&\varphi_{BD}= \frac{H}{\sqrt{2 k^3}} ( -k \eta + i ) e^{-ik\eta}\\
&A_+ = \frac{ e^{i(\pi/4 - b k/H)}}{\sqrt{1 - e^{- 2 \pi \alpha s k/H}}}\\
&A_- = \frac{ e^{ -\pi \alpha s k/H}e^{i(\pi/4 - b k/H)}}{\sqrt{1 - e^{- 2 \pi \alpha s k/H}}}
\end{split} \label{quantumstate}
\end{equation}
where $\eta$ is the usual conformal time in the de Sitter phase, $s$ is the planarity parameter defined in the Introduction and $\alpha=\frac{2^{2/3}}{3}, b=\frac{2^{2/3}\sqrt{\pi} \Gamma(1/3)}{3\Gamma(5/6)}$ are numerical factors \footnote{Note that in the limit where one takes $s \to 0$ holding $k$ fixed, the particle production in this model is not exponentially suppressed even at large $k$. This is related to the breakdown of the WKB condition for the scalar modes in the early Kasner phase around the point $s=0$ for a generic wavenumber. We refer the reader to section IV of \cite{Kim:2012bv} for further details.}.

In \cite{Anderson:2005hi}, a manifestly covariant renormalization scheme was used to obtain explicit formulae for independent components of the renormalized energy-momentum tensor for a scalar field in a FRW background. For the particular case of a de Sitter space, the energy density of a massless, minimally coupled scalar field in a generic excited state of the form given in the first line of equation (\ref{quantumstate}), is 
\begin{equation}
\begin{split}
&\epsilon= \epsilon_{BD} +I_1 +I_2\\
&I_1 \equiv \frac{1}{2 \pi^2} \int_0^{\infty} \,\, dk k^2 \left( |A_-|^2  |\dot{\varphi}_{BD}|^2 + {\rm Re}[ A_+ A_-^* ( \dot{\varphi}_{BD})^2] \right)\\
&I_2 \equiv \frac{1}{2 \pi^2 a^2} \int_0^{\infty} \,\, dk k^4 \left( |A_-|^2  |\varphi_{BD}|^2 + {\rm Re}[ A_+ A_-^* ( \varphi_{BD})^2] \right) 
\end{split}
\end{equation}
where $I_1$ and $I_2$ are depend on the particular quantum state through the coefficients $A_{\pm}$ and $a$ is the scale factor for the isotropic phase of inflation. 

One can now evaluate the above integrals from the formula of $A_{\pm}$ above in the limit $s \to 0$
\begin{equation}
\begin{split}
I_1&= \frac{1}{2 \pi^2}  \int_0^{\infty} \,\, dk k^2 \frac{k}{2 a^4} \left( \frac{ e^{ -2 \pi \alpha s k/H}}{1 - e^{- 2 \pi \alpha s k/H}} - \frac{ e^{ - \pi \alpha s k/H}}{1 - e^{- 2 \pi \alpha s k/H}} \cos\left( -2 k \eta + \frac{\pi}{2}- \frac{2 b k }{H}\right) \right) \\ 
 & =   \frac{ H^4}{960 a^4 \pi^2 \alpha^4 s^4} - \frac{i H^4}{128 a^4 \pi^6 \alpha^4 s^4}\left[\psi^{(3)}(\frac{1}{2}-\frac{i(2\eta H+2b)}{2\pi \alpha s})-\psi^{(3)}(\frac{1}{2}+\frac{i(2\eta H+2b)}{2\pi \alpha s})\right]\\
& \approx  \frac{ H^4}{960 a^4 \pi^2 \alpha^4 s^4} + \cdots
 \end{split}
\end{equation}

\begin{equation}
\begin{split}
 I_2&= \frac{1}{2 \pi^2 a^2}  \int_0^{\infty} \,\, dk k^4 \Big [\frac{ e^{ -2 \pi \alpha s k/H}}{1 - e^{- 2 \pi \alpha s k/H}} \frac{H^2}{2 k^3} (\frac{k^2}{a^2 H^2} +1) \\
 &+ \frac{H^2}{2 k^3}\frac{ e^{ - \pi \alpha s k/H}}{1 - e^{- 2 \pi \alpha s k/H}}\Big((\frac{k^2}{a^2 H^2} -1) \cos(- 2 k \eta + \frac{\pi}{2}- \frac{2 b k }{H})-\frac{2k}{aH}\sin(-2 k \eta + \frac{\pi}{2}- \frac{2 b k }{H})\Big)\Big]\\
 &= \frac{ H^4}{960 a^4 \pi^2 \alpha^4 s^4} +\frac{ H^4}{96 a^2 \pi^2 \alpha^2 s^2}  +\frac{iH^4}{32 a^2 \pi^4 \alpha^2 s^2}\left(\psi^{(1)}(\frac{1}{2}+\frac{i(-2\eta H-2b)}{2\pi \alpha s})-\psi^{(1)}(\frac{1}{2}-\frac{i(-2\eta H-2b)}{2\pi \alpha s})\right)\\
 & -\frac{iH^4}{128 a^4 \pi^6 \alpha^4 s^4}\left(\psi^{(3)}(\frac{1}{2}+\frac{i(-2\eta H-2b)}{2\pi \alpha s})-\psi^{(3)}(\frac{1}{2}-\frac{i(-2\eta H-2b)}{2\pi \alpha s})\right)\\
 &-\frac{H^4}{32 a^3 \pi^5 \alpha^3 s^3}\left(\psi^{(2)}(\frac{1}{2}-\frac{i(-2\eta H-2b)}{2\pi \alpha s})-\psi^{(2)}(\frac{1}{2}+\frac{i(-2\eta H-2b)}{2\pi \alpha s})\right)\\
 &\approx \frac{ H^4}{960 a^4 \pi^2 \alpha^4 s^4} +\cdots
 \end{split}
\end{equation}
where $\psi^{(m)}(z)$ are polygamma functions of order $n$.\\
 Since $\psi^{(m)}(z)=(-1)^{m+1} m! \sum^{\infty}_{k=0} \frac{1}{(k+z)^{m+1}}$ for any $m>0$ and any $z$ not equal to any negative integer, we have in the limit $s \to 0$
$$\psi^{(m)}(\frac{1}{2}+ 1/s) \sim s^{m+1} $$
which implies that all terms containing the polygamma functions are sub-leading in the limit $s \to 0$.\\
Therefore small back reaction will imply
\begin{equation}
\begin{split}
&10^{-4} \left(\frac{H^4}{a^4 s^4}\right) \ll H^2 M^2_p\\
&\implies s \gg  \frac{10^{-1}}{a} (\frac{H}{M_p})^{1/2} \implies s \gg s_{0} = \frac{10^{-1}}{a_{min}} (\frac{H}{M_p})^{1/2}
\end{split}
\end{equation}
It is reasonable to take $a_{min}$ as the scale factor close to the time of matching ($t_{*}$) of the WKB solution with the De Sitter solution
$$a_{min}=e^{\rho(t_{*})}=\left(\sqrt{\frac{k_{obs}}{H}}\right)_{min}$$
The observable limit of wavenumbers is given by 
\begin{equation}
 e^{N-64} \left( \frac{T_R}{10^{14} \mbox{GeV}} \right) \left( \frac{10^{16}\mbox{GeV}}{V^{1/4}}\right)^2 < \frac{k_{obs}}{H} < e^{N-55} \left( \frac{T_R}{10^{14} \mbox{GeV}} \right) \left( \frac{10^{16}\mbox{GeV}}{V^{1/4}}\right)^2 \label{limk/H}
 \end{equation}

The bound on $s$ from the back reaction consideration is therefore 
\begin{equation}
\boxed{s \gg s_0 = 10^{-1} (\frac{H}{M_p})^{1/2}\sqrt{\frac{H}{k_{obs}}} \implies \frac{\pi \alpha s k_{obs}}{H} \gg (\frac{H}{M_p})^{1/2}\sqrt{\frac{k_{obs}}{H}} } \label{brconstraint}
\end{equation}

A certain bound on $s$ also follows from the validity of the WKB solution and the matching procedure in the regime of planar modes (see section IV.B of \cite{Kim:2012bv}) namely 
\begin{equation}
\left( \frac{H}{k}\right)^{3/5} \gg s \gg \left( \frac{H}{k}\right)^3
\end{equation}
which also implies that $k/H >1$.

For $k_{obs} \gg H$, the bound from back reaction appears to be stronger than the bound obtained from the validity of WKB approximation. For example, for $\frac{k_{obs}}{H}\sim O(10^4)$, we have $s^{BR}_0 \sim 10^{-6}$ as opposed to $s^{WKB}_0 \sim 10^{-12}$. On the other hand, for $k_{obs} \sim 10 H$, the WKB bound is the stronger of the two. As an example, $s^{BR}_0 \sim 10^{-3.5}$ while $s^{WKB}_0 \sim 10^{-3}$. 

For the rest of the paper, we will focus on the range of momenta $k_{obs} \sim 10^2 H$, such that the lower bound on $\frac{\pi \alpha s k}{H}$ is $(\frac{\pi \alpha s k}{H})_0 \sim10^{-2}$. Note that this implicitly fine-tunes the number of e-foldings to $N \gtrsim 64$, as one can see from \eqref{limk/H}. For $N \gg 64$, although our computation holds, the back reaction constraint found in \eqref{brconstraint} will dictate that $\frac{\pi \alpha s k_{obs}}{H} \gg O(1)$, which will lead to exponential suppression of any signature of anisotropy. If $N < 64$, on the other hand, the WKB condition above will be violated. Therefore, $N \gtrsim 64$ is the regime where we expect to see non-trivial effects of primordial anisotropy.

\section{Standard Action: Physical Observables}\label{stanact}

\subsection{Power Spectrum}
The power spectrum for the planar modes of the scalar perturbation was computed in \cite{Dey:2012qp} and can be written explicitly as a function of the planarity parameter $s$
\begin{equation}
\boxed{P(k)=P(k)_0\left(\coth{(\pi \alpha s k/H)}- \frac{\sin{(2bk/H)}}{\sinh{(\pi \alpha s k/H)}}\right)}
\label{PS}
\end{equation}
where $P(k)_0$ is the power spectrum for standard inflation \cite{Dey:2012qp}. For $\pi \alpha s k/H \sim 10^{-2}$, one can simplify the formula for the power spectrum by expanding the hyperbolic functions with the argument $\pi \alpha s k/H$ held as a small parameter. Therefore, we have
\begin{equation}
\boxed{P(k)=P(k)_0 \left( \frac{1-\sin{(2bk/H)}}{\pi \alpha s k/H} \right)} \label{psplanar}
\end{equation}
\subsection{Non-Gaussianity}
In this section, we derive expressions for the $f_{\rm NL}$ parameters for planar modes and study these in the regime where $\pi \alpha s k/H \sim 10^{-2}$. It is convenient to set $M_p=1$ for this computation and reinstate the appropriate factors of $M_p$ in the final formula for $f_{NL}$ using dimensional analysis.

The 3-point correlation function for the planar modes can be computed in the ``in-in" formalism using the interaction Hamiltonian $$ {\cal H}_I=-\int d^3x \, d\eta \, e^{3\rho} \left(\frac{\dot{\phi}}{\dot{\rho}}\right)^4  \dot{\rho} \,\zeta_c^{'2}\partial^{-2}\zeta'_c \label{intHam}$$ where $\zeta_c$ is related to $\zeta$ by a local (in time) non-linear field redefinition
$$\zeta=\zeta_c +\frac{\ddot{\phi}}{2\dot{\phi}\dot{\rho}}\zeta_c^2 + \frac{\dot{\phi}^2}{8\dot{\rho}^2}\zeta_c^2 + \frac{\dot{\phi}^2}{4\dot{\rho}^2}\partial^{-2}(\zeta_c\partial^2\zeta_c)$$ Evidently, this redefinition does not change the quadratic action which implies that $\zeta_c$ and $\zeta$ have the same equation of motion and hence the same classical solution.\\

Therefore, the appropriate quantum state to be used in the computation of the 3-point function is the one specified in equation (\ref{quantumstate}). One can now use the ``in-in'' formalism to compute the tree-level contributions to the 3-point correlation function of scalar perturbations. Since there is only one kind of interaction vertex, we have only two distinct Feynman diagrams at the tree-level -- one with a ``right'' vertex and the other one with a ``left''.  Recall that for any observable $Q(t)$, $\left\langle Q(t) \right\rangle_{in-in} =\left\langle [\bar{T}\exp(i\int^t_{t_0}H_I(t)dt)] Q^I(t)[T\exp(-i\int^t_{t_0}H_I(t)dt)] \right\rangle$, where $T$ and $\bar{T}$ denotes the time-ordered and the anti-time-ordered product of operators. One needs to distinguish between vertices arising out of the time-ordered product from those coming from the anti-time-ordered product and we refer to them as ``right'' and ``left'' vertices respectively.

The three-point correlation function in this formalism is
\begin{equation}
\left\langle \zeta_c({\bf k_1},\eta)\zeta_c({\bf k_2},\eta)\zeta_c({\bf k_3},\eta)\right\rangle =(2\pi)^3 \delta^{(3)}({\bf k_1}+ {\bf k_2}+{\bf k_3})\left[A^{R}({\bf k_1},{\bf k_2}, {\bf k_3},\eta) +c.c. \right]
\end{equation}
where the``right" amplitude, at late times, can be written as
\begin{equation}
A^R({\bf k_1},{\bf k_2},{\bf k_3})=-i\frac{\sum_{i<j}k^2_ik^2_j}{\dot{\phi}^2}
\int^0_{\eta_0}d\eta'\sum_{{\xi_i}=\pm1}\prod^3_{i=1}e^{-i(\xi_ik_i)\eta'}  F_{\xi_i}(k_i)
\end{equation}
where the sum extends over all 8 possible linear combinations $\sum \xi_ik_i$ and
$F_{\xi_i=-1}(k_i)=\frac{H^2}{2k^3_i}(|A^{i}_{+}|^2-A^{i}_{-}{A^{i}_{+}}^{\ast})$ and $F_{\xi_i=1}(k_i)=\frac{H^2}{2k^3_i}(|A^{i}_{-}|^{2}-A^{i}_{+}{A^{i}_{-}}^{\ast})$.\\

Therefore, on completing the $\eta'$ integration, we have
\begin{equation}
\boxed{A^R({\bf k_1},{\bf k_2},{\bf k_3})=-\frac{\sum_{i<j}k^2_ik^2_j}{\dot{\phi}^2}
\sum_{{\xi_i}=\pm1}(\prod^3_{i=1} F_{\xi_i}(k_i))\frac{1}{\sum_i\xi_ik_i}\left(1-e^{-i \eta_0\sum_i\xi_ik_i}\right)}  \label{aRf2}
\end{equation}
Let us define $\tilde{F_{\xi_i}}(k_i)=\frac{2k^3_i}{H^2} F_{\xi_i}(k_i)$, which can be written as explicit functions of wavenumber as follows:
\begin{eqnarray}
\tilde{F}_{-}(k_i)&=&\frac{(1- e^{-\pi (\alpha s k_i/H)}e^{-i(\pi/2 - 2 b k_i/H)})}{1-e^{-2\pi (\alpha s k_i/H)}}=\frac{e^{\pi \alpha s k_i/H}+i e^{i2bk_i/H}}{2\sinh{(\pi \alpha s k/H)} }\approx \frac{1+ie^{i2bk_i/H}}{2\pi \alpha s k/H}\\ \nonumber \\
\tilde{F}_{+}(k_i)&=&-e^{-\pi \alpha s k_i/H}e^{i(\pi/2 - 2 b k_i/H)} \tilde{F}_{-}(k_i) \approx -i e^{- 2i b k_i/H} \frac{1+ie^{i2bk_i/H}}{2\pi \alpha s k/H}
\end{eqnarray}
The amplitude computed above is enhanced for the flattened triangle configuration as well as the squeezed triangle configuration and are analyzed below.\\ \\
\noindent \textbf{Flattened Triangle Configuration}\\
 In this case, the enhancement appears when the wave-numbers satisfy $\sum \xi_{i} k_i=0$, so that the exponent of the exponential term in equation (\ref{aRf2}) vanishes.\\
We choose $k_2 \approx k_3 \approx k_1/2 $, setting $k_1=k_2 + k_3$. For this choice, the set $\{\xi_i\}$ contributing to the enhanced bispectrum can have values $(1,-1,-1)$ and $(-1,1,1)$. Therefore,
\begin{equation}
A^R({\bf k},{\bf k},{\bf k})=\frac{3 H^6}{8 k^5 \dot{\phi}^2_e} \tilde{F}^3_{-}({\bf k}) \eta_0 (e^{-\pi \alpha s k/H}e^{-i2 b k/H}- ie^{-2\pi \alpha s k/H}e^{-i4 b k/H})
\end{equation}
Using the definition of $f^{\rm flat}_{\rm NL}: f^{\rm flat}_{\rm NL}=(A^R({\bf k},{\bf k},{\bf k}) + c.c.)/P({\bf k})^2$, we have
\begin{equation}
f^{\rm flat}_{\rm NL}=\frac{\dot{\phi}^2_e}{M^2_p H^2} \left(\frac{\cos{(2bk/H)}\cosh{(\pi \alpha s k/H)} -\sin{(4bk/H)}}{\left(\cosh{(\pi \alpha s k/H)}-\sin{(2bk/H)}\right)^2}\right) \label{genfNL_flat}
\end{equation}
where we have taken off the $k\eta_0$ factor in the estimation of the physical $f_{\rm NL}$ owing to the 2D projection of the naive formula for $f_{\rm NL}$ as discussed in \cite{Holman:2007na}. Note that we have plugged in the appropriate factor of $M_p$ in this formula.\\
In the small $s$ limit such that $\pi \alpha s k/H \ll 1$, we have
\begin{equation}
\boxed{f^{\rm flat}_{\rm NL}\approx  \frac{\dot{\phi}^2_e}{M^2_p H^2}\left(\frac{\cos{(2bk/H)} -\sin{(4bk/H)}}{(1-\sin{(2bk/H)})^2}\right)}
\end{equation}
Note that $f^{\rm flat}_{\rm NL}$ is completely independent of the planarity parameter in this limit and is suppressed by the slow-roll parameter, namely
\begin{equation}
\boxed{|f^{\rm flat}_{\rm NL}| \approx  \frac{\dot{\phi}^2_e}{M^2_p H^2}}
\end{equation}
In the limit $\pi \alpha s k/H \gg 1$, we have
\begin{equation}
\boxed{|f^{\rm flat}_{\rm NL}| \approx  \frac{\dot{\phi}^2_e}{M^2_p H^2} e^{-\pi \alpha s k/H}}
\end{equation}
Note that the $f_{\rm NL}$ is further suppressed in this case by the exponential factor.\\

\noindent \textbf{Squeezed Triangle Configuration}\\

For the squeezed triangle configuration, $k_3 \ll k_1 \approx k_2 \sim k$, and the corresponding amplitude is given as
\begin{equation}
A^R({\bf k},{\bf k},{\bf k_3}) \approx -\frac{k^4}{\dot{\phi}^2}\frac{H^3}{8k^6 k^3_3} \sum_{{\xi_i}=\pm1}(\prod^3_{i=1} \tilde{F_{\xi_i}}(k_i))\frac{1}{\sum_i\xi_ik_i}\left(1-e^{-i \eta_0\sum_i\xi_ik_i}\right) 
\end{equation}
where the sum is over the set   $\xi_i : (1,-1,-1),(-1,1,1),(1,-1,1),(-1,1,-1)$.\\
Let $s_3$ denote the planarity parameter for the smaller vector ${\bf k_3}$, while $s$ denotes the planarity parameter of the vector ${\bf k}$. The formula for $f^{\rm sqzd}_{\rm NL}$ for generic values of $s$ and $s_3$ has been derived in the appendix - we only quote the final result here
\begin{equation}
\begin{split}
&f^{\rm sqzd}_{\rm NL}=\left(\frac{\dot{\phi}^2_e}{M^2_p H^2}\right) \left(\frac{k}{k_3}\right) \frac{F(k,k_3, \eta_0,s_3)}{\sinh{\pi \alpha s k/H} (\cosh{\pi \alpha s k/H}-\sin{2bk/H})(\cosh{\pi \alpha s_3 k_3/H}-\sin{2bk_3/H})}\\
&F(k,k_3, \eta_0,s_3)=4\Big[2\sinh{\pi \alpha s_3 k_3/H}\sin{2bk/H}\cos{\pi \alpha s k/H}\\
&- \sin{(2bk/H+\eta_0k_3)}\sinh{(\pi \alpha s k/H+\pi \alpha s_3 k_3/H)} + \sin{(2bk/H-\eta_0k_3)}\sinh{(\pi \alpha s k/H-\pi \alpha s_3 k_3/H)} \\
&-2\cos{2bk_3/H}\cos{2bk/H}\sinh{\pi \alpha s k/H}+2\cos{(2bk_3/H- \eta_0 k_3)}\cos{2bk/H}\sinh{\pi \alpha s k/H}\Big]
\end{split} \label{genfNL_sqzd}
\end{equation}

First, assume that $s \sim s_3$, then $\pi \alpha s_3 k_3/H \ll \pi \alpha s k/H$, since $k_3 \ll k$. For the WKB condition to be satisfied one needs to take the smaller wavenumber $k_3 \sim 10 H$. If we demand that both $\pi \alpha s_3 k_3/H $ and $\pi \alpha s k/H$ are smaller than $O(1)$, then the ratio $\frac{k}{k_3}$ is constrained to be of $O(10)$ at most. In this regime, we have
\begin{equation}
\begin{split}
\boxed{f^{\rm sqzd}_{\rm NL}\approx \left(\frac{\dot{\phi}^2_e}{M^2_p H^2}\right) \left(\frac{k}{k_3}\right) \frac{2\cos{\frac{2bk}{H}}\left(\cos{\eta_0k_3}+\cos{(\frac{2bk_3}{H}-\eta_0k_3)}-\cos{\frac{2bk_3}{H}}\right)}{(1-\sin{2bk/H})(1-\sin{2bk_3/H})}}
\end{split}
\label{squeezed1}
\end{equation}

One can estimate the order of magnitude of the $f_{\rm NL}$ parameter for a generic wavenumber $k$ for a small planarity parameter $s$
\begin{equation}
\boxed{|f^{\rm sqzd}_{\rm NL}| \approx  \left(\frac{\dot{\phi}^2_e}{M^2_p H^2}\right)\left(\frac{k}{k_3}\right)} \label{planarsqpure1}
\end{equation}
Since the slow-roll parameter $\epsilon \sim \frac{\dot{\phi}^2_e}{M^2_p H^2} \sim 10^{-2}$ and $\left(\frac{k}{k_3}\right)_{max} \sim 10$ , $f^{\rm sqzd}_{\rm NL}$ could at best be of $O(10^{-1})$.\\
One can also consider the situation where $\pi \alpha s_3 k_3/H $ is much smaller than $O(1)$, but $\pi \alpha s k/H $ is larger or of order unity. In that case, one can have a larger ratio for $\frac{k}{k_3}$. Explicitly, the order of magnitude estimate is 
\begin{equation}
\boxed{|f^{\rm sqzd}_{\rm NL}| \approx  \left(\frac{\dot{\phi}^2_e}{M^2_p H^2}\right)\left(\frac{k}{k_3}\right)e^{-\pi \alpha s k/H}} \label{planarsqpure2}
\end{equation}
This shows that in spite of having a larger enhancement from the $\frac{k}{k_3}$ factor, the $f_{\rm NL}$ parameter now has an additional suppression from the exponential term which constrains the maximum value of $f^{\rm sqzd}_{\rm NL}$ to again be of $O(10^{-1})$.\\

Note that in either case ($\pi \alpha s k/H$ being small or large), the squeezed triangle  configuration leads to larger non-Gaussianity  compared to the flattened triangle case
\begin{equation}
\frac{|f^{\rm sqzd}_{\rm NL}|}{|f^{\rm flat}_{\rm NL}|} \approx \frac{k}{k_3} > 1
\end{equation}


\section{Higher Derivative Operator: Non-Gaussianity}\label{hdact}
Consider a single scalar field model of inflation with a dimension-8 operator
\begin{equation} 
S=\frac{M^2_p}{2}\int d^4x \sqrt{-g} R + \int d^4x \sqrt{-g}\left(-\frac{1}{2}g^{\mu\nu}\partial_{\mu}\phi\partial_{\nu}\phi -V(\phi)\right)+\int d^4x\sqrt{-g} \frac{\lambda}{8M^4}(\nabla\phi)^4
\end{equation}
where $M$ is the cut-off scale for the effective theory of inflation and $\lambda$ is a dimensionless parameter. Since the dimension 8 operator respects the shift symmetry, one can see that it does not spoil the slow-roll conditions.\\
 \indent   Around a homogeneous background, one can derive the quadratic lagrangian for the curvature perturbation $\zeta$, which now includes contribution from the dimension 8 operator as well. However, the correction to the mode functions is $O(\lambda)$ and since the interaction Hamiltonian itself is $O(\lambda)$, we can safely neglect such corrections to the mode function in the computation of the three-point correlation functions. The power spectrum for scalar perturbations in the theory  therefore receives correction only at $O(\lambda)$.

In computing the 3-point correlation functions, we again set $M_p=1$ for convenience and reinstate appropriate factors of $M_p$ in the final formula for $f_{NL}$ using dimensional considerations. Expanding the classical action to third order in curvature perturbation, we obtain the following interaction Hamiltonian from the dimension 8 operator \cite{Holman:2007na,Creminelli:2004yq}
\begin{equation}
{\cal H}_I=-\int d^3x a(\eta)\frac{\lambda \dot{\phi}^4}{2H^3M^4}\zeta'(\zeta'^2-(\partial_i\zeta)^2)  \label{asf}
\end{equation}

From ($\ref{asf}$), one can directly compute the 3-point correlation function
\begin{eqnarray}
A_R({\bf k_1},{\bf k_2},{\bf k_3})&=&i\int^{0}_{\eta_0}d\eta e^{\rho(\eta)} \frac{\lambda \dot{\phi}^4}{\dot{\rho}^3 M^4} [\prod^{3}_{i=1}\partial_{\eta}G_{{\bf k_i}}(0,\eta)\times (3!) \nonumber \\ &+& ((\vec{k_1}.\vec{k_2}) G_{{\bf k_1}}(0,\eta)G_{{\bf k_2}}(0,\eta)\partial_{\eta}G_{{\bf k_3}}(0,\eta) +\mbox{perms})\times (2!) ]  \label{aqf}\nonumber\\&:=&A^{(1)}_R({\bf k_1},{\bf k_2},{\bf k_3})+A^{(2)}_R({\bf k_1},{\bf k_2},{\bf k_3})
\end{eqnarray}
where the factors of $3!$ and $2!$ are the respective combinatorial factors for the two vertices and $A^{(1)}_R({\bf k_1},{\bf k_2},{\bf k_3}),A^{(2)}_R({\bf k_1},{\bf k_2},{\bf k_3})$ denote the contributions of the two vertices 
to the ``right"  amplitude.\\
The functions $G_{{\bf k_i}}(0,\eta)$ and $\partial_{\eta}G_{{\bf k_i}}(0,\eta)$ are given in terms of the coefficients $A_{\pm}({\bf k_i})$ as follows:
\begin{eqnarray}
G_{{\bf k_i}}(0,\eta)& = & \frac{\dot{\rho}^2}{\dot{\phi}^2} \frac{\dot{\rho}^2}{2k^3_i}\left( |A_{+}|^2(1-ik_i\eta)e^{ik_i\eta}+|A_{-}|^2(1+ik_i\eta)e^{-ik_i\eta}  \right. \nonumber\\
& & - \left.  A_{+}A_{-}^{\ast}(1+ik_i\eta)e^{-ik_i\eta}-A_{-}A_{+}^{\ast}(1-ik_i\eta)e^{ik_i\eta}\right)
\end{eqnarray}
\begin{equation}
\partial_{\eta}G_{{\bf k_i}}(0,\eta)=-\frac{\dot{\rho}^2}{\dot{\phi}^2}{k_i^2\eta}\frac{\dot{\rho}^2}{2k^3_i}\left[(|A_{+}|^2-A_{-}A_{+}^{\ast})e^{ik_i\eta}+(|A_{-}|^2-A_{+}A_{-}^{\ast})e^{-ik_i\eta}\right]
\end{equation}
In the last equation $\eta \approx -\frac{1}{\dot{\rho}\exp{\rho(\eta)}}$, which is valid in the de Sitter phase of expansion. Since we have chosen $\eta_0 \approx \eta_{iso}$, this is a good approximation for $\eta \in [\eta_0,0]$.\\

 The contributions of the two vertices specified above can be re-written in terms of $\tilde{F}_{\pm}$ as
\begin{equation}
A^{(1)}_R({\bf k_1},{\bf k_2},{\bf k_3})=-i\int^{0}_{\eta_0}d\eta e^{\rho(\eta)} \frac{\lambda \dot{\phi}^4}{\dot{\rho}^3 M^4} (\frac{\dot{\rho}^2\eta}{\dot{\phi}^2})^3\prod^{3}_{i=1}\frac{\dot{\rho}^2}{2k_i} \prod^{3}_{i=1}\left[\tilde{F}_{+}({\bf k_i})e^{-ik_i\eta}+\tilde{F}_{-}({\bf k_i})e^{ik_i\eta}\right]\times (3!) \label{rightamp1}
\end{equation}
\begin{equation}
\begin{split}
A^{(2)}_R({\bf k_1},{\bf k_2},{\bf k_3})=&-i\int^{0}_{\eta_0}d\eta e^{\rho(\eta)} \frac{\lambda \dot{\phi}^4\eta}{\dot{\rho}^3 M^4} \prod^{3}_{i=1}\frac{\dot{\rho}^4}{2\dot{\phi}^2 k^3_i} [(\vec{k_1}.\vec{k_2})k^2_3\prod^{2}_{i=1}\left((1+ik_i\eta)\tilde{F}_{+}({\bf k_i})e^{-ik_i\eta}+(1-ik_i\eta)\tilde{F}_{-}({\bf k_i})e^{ik_i\eta}\right)\\
&\times\left(\tilde{F}_{+}({\bf k_3})e^{-ik_3\eta}+\tilde{F}_{-}({\bf k_3})e^{ik_3\eta}\right)+\mbox{perms.}]\times (2!) 
\end{split} \label{rightamp2}
\end{equation}

\noindent \textbf{Flattened Triangle Configuration}\\
We choose $k_3 \approx k_2 \approx k_1/2 \approx k$, which sets $k_1=k_2+k_3$, such that the integrands appearing in the "right" amplitudes $A^{(1)}_R$ and $A^{(2)}_R$ in Eq.~(\ref{rightamp1}) and Eq.~(\ref{rightamp2}) respectively have leading order contributions from the following configurations : $\{+,-,-\}$ and $\{-,+,+\}$. The details of the computation of the three-point correlation function for the flattened triangle configuration can be found in the appendix. The final form of $f^{\rm flat}_{\rm NL}$ for generic values of the planarity parameter is 
\begin{equation}
f^{\rm flat}_{\rm NL}=- \frac{3 \lambda \dot{\phi}^2_e}{16 M^4} k\eta_0 \left(\frac{\cos{(4bk/H)} -\sin{(4bk/H)}}{\left(\cosh{(\pi \alpha s k/H)}-\sin{(2bk/H)}\right)^2}\right) \label{hdfNL_flat}
\end{equation}
In the limit  $\pi \alpha s k/H \ll 1$, we have,
\begin{equation}
\boxed{ f^{\rm flat}_{\rm NL} \approx - \frac{3 \lambda \dot{\phi}^2_e}{16 M^4} k\eta_0 \left(\frac{\cos{(4bk/H)} -\sin{(4bk/H)}}{\left(1-\sin{(2bk/H)}\right)^2}\right)}
\end{equation}
For a generic wavenumber $k$, one can estimate the magnitude of the $f_{\rm NL}$ parameter
\begin{equation}
\boxed{|f^{\rm flat}_{\rm NL}| \approx   \frac{ \lambda \dot{\phi}^2_e}{M^4} k\eta_0 \leq \lambda \epsilon \left(\frac{H}{M_p}\right) \left(\frac{M_p}{M}\right)^3}
\end{equation}
where in the last step we have used $\epsilon=\frac{\dot{\phi}^2_e}{M^2_p H^2}$ and $|k\eta_0| \leq \frac{M}{H}$.\\
In the limit $\pi \alpha s k/H \gg 1$, the $f_{\rm NL}$ parameter is exponentially suppressed
\begin{equation}
\boxed{|f^{\rm flat}_{\rm NL}| \approx   \frac{ \lambda \dot{\phi}^2_e}{M^4} k\eta_0 e^{-2\pi \alpha s k/H}}
\end{equation}

\noindent \textbf{Squeezed Triangle Configuration}\\
For the squeezed triangle configuration, $k_3 \ll k_1 \approx k_2 \sim k$, and therefore the configurations that contribute to the integrands  of the "right" amplitudes $A^{(1)}_R$ and $A^{(2)}_R$ at the leading order are $\{+,-,-\},\{-,+,+\},\{+,-,+\},\{-,+,-\}$. The details of the computation of the three-point correlation function and the $f_{\rm NL}$ parameter for generic values of the planarity parameter are worked out in the appendix. Here, we only write down the final result
\begin{equation}
\begin{split}
f^{\rm sqzd}_{\rm NL}= \frac{ 4 \lambda \dot{\phi}^2_e}{M^4}\left(\frac{k}{k_3}\right) \left(\frac{\sin{\frac{k_3\eta_0}{2}}\left(\sinh{\frac{\pi \alpha s k}{H}}U(k,k_3,\eta_0)-\sinh{\frac{\pi \alpha s k_3}{H}}V(k,k_3,\eta_0)\right)}{\sinh{\pi \alpha s k/H}\left(\cosh{(\pi \alpha s k/H)}-\sin{(2bk/H)}\right)\left(\cosh{(\pi \alpha s k_3/H)}-\sin{(2bk_3/H)}\right)}\right)
\end{split}
\label{hdfNL_sqzd}
\end{equation}
where $U$ and $V$ are functions (independent of $s$) written explicitly in the appendix.\\
In the limit  $\pi \alpha s k/H \ll 1$, one can therefore write $f^{\rm sqzd}_{\rm NL}$ as
\begin{equation}
\begin{split}
\boxed{f^{\rm sqzd}_{\rm NL} \approx \frac{ 4 \lambda \dot{\phi}^2_e}{M^4}\left(\frac{k}{k_3}\right)\left( \frac{\sin{\frac{k_3\eta_0}{2}}U(k,k_3,\eta_0)}{\left(1-\sin{(2bk/H)}\right)\left(1-\sin{(2bk_3/H)}\right)}\right)} \label{squeezed2}
\end{split}
\end{equation}
For a generic wavenumber $k$, one can estimate the magnitude of the $f_{\rm NL}$ parameter
\begin{equation}
\begin{split}
\boxed{f^{\rm sqzd}_{\rm NL} \approx \frac{  \lambda \dot{\phi}^2_e}{M^4}\left(\frac{k}{k_3}\right)}
\end{split}
\end{equation}

Note that $f^{\rm flat}_{\rm NL}$ is larger in magnitude compared to $f^{\rm sqzd}_{\rm NL}$, as opposed to the case of standard inflationary action
\begin{equation}
\boxed{\frac{|f^{\rm flat}_{\rm NL}|}{|f^{\rm sqzd}_{\rm NL}|} \approx k_3 \eta_0 >1}
\end{equation}
In the limit $\pi \alpha s k/H \gg 1$ and $\pi \alpha s k_3/H \ll 1$, the $f_{\rm NL}$ parameter is exponentially suppressed
\begin{equation}
\boxed{|f^{\rm sqzd}_{\rm NL}| \approx   \frac{ \lambda \dot{\phi}^2_e}{M^4} \frac{k}{k_3} e^{-\pi \alpha s k/H}}
\end{equation}

\section{Conclusion and Discussion}\label{mainres}

To conclude, we begin by summarizing the main results of our paper. We review the field theory results for  the case of single (scalar) field inflation with standard action and higher derivative operators, respectively, in a Kasner-de Sitter background. In the remainder of this section, we move on to a computation of the signal-to-noise ratio of the anisotropy of the CMB power spectrum and discuss the detectability of our model in Cosmic Microwave Background (CMB) and large scale structure data.\\

\subsection{Main Results : Power Spectrum and Non-Gaussianities}
We consider a single scalar inflaton driving an axially symmetric Bianchi I geometry,
\begin{equation} 
\begin{split}
&S=\frac{M^2_p}{2}\int d^4x \sqrt{-g} R + \int d^4x \sqrt{-g}\left(-\frac{1}{2}g^{\mu\nu}\partial_{\mu}\phi\partial_{\nu}\phi -V(\phi)\right),\\
&ds^2= -dt^2 + e^{2\rho} dx^2 + e^{2\beta} (dy^2 +dz^2)
\end{split}
\end{equation}
\indent The geometry in question is of Kasner type at early times and isotropizes to a de Sitter phase at late times. Scalar cosmological perturbations in this theory have a particularly interesting regime where the wave-number vector of a given mode lies almost entirely in a plane. Such modes may be labelled by the magnitude of the wave-number vector $k$ and the planarity parameter $s=\frac{k_x}{k}$, where the vector ${\bf k}$ is understood to lie almost completely on the $y-z$ plane ($k_x \ll k$). As discussed in \cite{Dey:2011mj,Kim:2011pt,Dey:2012qp}, the information of early time anisotropy manifests itself as an excited state (a Bogoliubov-transformed Bunch-Davies state) for the scalar perturbations in the late time de Sitter phase. This excited state may be explicitly calculated by matching the early time WKB solution with the late time de Sitter solution \cite{Dey:2011mj, Kim:2011pt, Dey:2012qp,Kim:2010wra}.\\
\indent One can therefore proceed to computing physical observables for the scalar perturbations at late times in the aforementioned excited state using the ``in-in" formalism. The power spectrum which follows from the two-point correlation function is given by
\begin{equation}
\begin{split}
&P(k)=P(k)_0\left(\coth{(\pi \alpha s k/H)}- \frac{\sin{(2bk/H)}}{\sinh{(\pi \alpha s k/H)}}\right) 
\end{split}
\end{equation}
where $P(k)_0$ is the power spectrum for standard inflation \cite{Dey:2012qp} and $\alpha=\frac{2^{2/3}}{3}$ and $b=\frac{2^{2/3}\sqrt{\pi} \Gamma(1/3)}{3\Gamma(5/6)}$ are numerical factors. In section \ref{br}, we performed a back reaction computation to obtain a lower bound for the parameter $s$. In addition, there are bounds coming from the requirement that the modes satisfy WKB approximation at early times. Together, we have
\begin{equation}
\begin{split}
&\frac{\pi \alpha s k_{obs}}{H} \gg (\frac{H}{M_p})^{1/2}\sqrt{\frac{k_{obs}}{H}} \; \; (\mbox{\bf Back Reaction})\\
&\left( \frac{H}{k_{obs}}\right)^{3/5} \gg s \gg \left( \frac{H}{k_{obs}}\right)^3 \; \; (\mbox{\bf WKB})
\end{split} \label{br1}
\end{equation}
The 3-point correlation functions of the scalar perturbations are enhanced in the flattened triangle and the squeezed triangle configurations. The $f_{\rm NL}$ parameters in the respective configurations for generic values of the planarity parameters can be found in section \ref{stanact} (see equation (\ref{genfNL_flat}) and equation (\ref{genfNL_sqzd}) for example). The general formulae for $f_{\rm NL}$ in the flattened and squeezed triangle configurations clearly show that there are singularities at certain discrete values of $k$, where one might get a large enhancement \footnote{This enhancement of $f_{NL}$ should not be confused with the enhancement of the 3-point correlation function, as the former arises precisely from the zeroes of the power spectrum.}. The effect of such singularities on observable non-Gaussianity will be a subject of future work, as discussed below. Here we focus on generic modes which are far away from any such singularity.\\ 
In the particular limit where $\frac{\pi \alpha s k}{H},\; \frac{\pi \alpha s_3 k_{3}}{H}  \ll 1$, we have
\begin{equation}
\begin{split}
& |f^{\rm flat}_{\rm NL}| \approx  \frac{\dot{\phi}^2_e}{M^2_p H^2}\\
& |f^{\rm sqzd}_{\rm NL}| \approx  \left(\frac{\dot{\phi}^2_e}{M^2_p H^2}\right)\left(\frac{k}{k_3}\right) \;\; (k \gg k_3)
\end{split}
\end{equation}
This limit can only be taken if $H \sim 10^{-5} M_p$ or lower and the wave-number is not too large, namely $k \sim 10H - 100H$. Therefore, 
the back reaction condition, given by the first line in equation (\ref{br1}), constrains the ratio $\frac{k}{k_3} \sim 10$, if the wave number vectors ${\bf k}$ and ${\bf k_3}$ are such that $\frac{\pi \alpha s k}{H},\; \frac{\pi \alpha s_3 k_{3}}{H}  \ll 1$. Therefore, for a generic wave number, $f^{\rm sqzd}_{\rm NL}$ can at best be of $O(10^{-1})$.\\
\noindent One can also consider the limit where $\frac{\pi \alpha s k}{H} \gg 1$ while $\frac{\pi \alpha s_3 k_{3}}{H}  \ll 1$. In this case, the ratio $\frac{k}{k_3}$ can be larger, but this enhancement is now compensated by an additional exponential suppression
\begin{equation}
\begin{split}
& |f^{\rm flat}_{\rm NL}| \approx  \frac{\dot{\phi}^2_e}{M^2_p H^2} e^{-\frac{\pi \alpha s k}{H}}\\
& |f^{\rm sqzd}_{\rm NL}| \approx  \left(\frac{\dot{\phi}^2_e}{M^2_p H^2}\right)\left(\frac{k}{k_3}\right)e^{-\frac{\pi \alpha s k}{H}} \;\; (k \gg k_3)
\end{split}
\label{ExpSup1}
\end{equation}
However, in either case, note that the squeezed configuration leads to larger non-Gaussianity compared to the flattened configuration
\begin{equation}
\frac{|f^{\rm sqzd}_{\rm NL}|}{|f^{\rm flat}_{\rm NL}|} \approx \frac{k}{k_3} > 1
\end{equation}
\\
Next, we have considered a single scalar inflaton in the same axially symmetric Bianchi I space-time, but with a dimension 8 operator
\begin{equation} 
S=\frac{M^2_p}{2}\int d^4x \sqrt{-g} R + \int d^4x \sqrt{-g}\left(-\frac{1}{2}g^{\mu\nu}\partial_{\mu}\phi\partial_{\nu}\phi -V(\phi)\right)+\int d^4x\sqrt{-g} \frac{\lambda}{8M^4}(\nabla\phi)^4 \label{Actionhd}
\end{equation}
where $M$ is the cut-off scale for the effective theory of inflation and $\lambda$ is a dimensionless parameter. The physics of this particular higher derivative operator has been discussed earlier in \cite{Holman:2007na, Creminelli:2003iq}. In this paper, we focused on the role of such an operator in anisotropic space-time. As explained in section \ref{hdact}, the power spectrum of scalar perturbations is only corrected at $O(\lambda)$ or higher. For the computation of non-Gaussianities to the first order in $\lambda$, such corrections may be ignored. Also, the condition on the planarity $s$ coming from back reaction is only corrected at $O(\lambda)$ due to the addition of the dimension 8 operator.\\
In this theory, we computed the 3-point correlation function of curvature perturbations at $O(\lambda)$ and noted that there are enhancements in the flattened triangle and the squeezed triangle configurations. The $f_{\rm NL}$ parameters in the respective configurations for generic values of the planarity parameters can be found in section \ref{hdact} (see Eq.~(\ref{hdfNL_flat}) and Eq.~(\ref{hdfNL_sqzd})). As in the case of single field inflation with standard action discussed before, we have restricted ourselves to modes which are away from the singularities.\\
 In the particular limit where $\frac{\pi \alpha s k}{H},\; \frac{\pi \alpha s_3 k_{3}}{H}  \ll 1$, we have
\begin{equation}
\begin{split}
& |f^{\rm flat}_{\rm NL}| \approx  \frac{ \lambda \dot{\phi}^2_e}{M^4} k\eta_0 \leq \lambda \epsilon \left(\frac{H}{M_p}\right) \left(\frac{M_p}{M}\right)^3 \\
& |f^{\rm sqzd}_{\rm NL}| \approx  \frac{  \lambda \dot{\phi}^2_e}{M^4}\left(\frac{k}{k_3}\right) \leq   \lambda \epsilon \left(\frac{H}{M_p}\right)^2 \left(\frac{M_p}{M}\right)^4 \;\; (k \gg k_3)
\end{split}
\end{equation}
In this limit, $f^{\rm flat}_{\rm NL}$ is larger in magnitude compared to $f^{\rm sqzd}_{\rm NL}$, as opposed to the case of standard inflationary action
\begin{equation}
\frac{|f^{\rm flat}_{\rm NL}|}{|f^{\rm sqzd}_{\rm NL}|} \approx k_3 \eta_0 >1
\end{equation}
For $\lambda \sim O(1)$ and $M \sim 10^{-2} M_p$, we have $|f^{\rm flat}_{\rm NL}| \sim O(10^{-1})$ while $|f^{\rm sqzd}_{\rm NL}| \sim O(10^{-4})$. For larger $f_{\rm NL}$, one needs to make the ratio $M/M_p$ smaller, in which case the higher dimensional operators (higher than dimension 8) cannot be ignored in the effective inflaton action.\\

In the limit where $\frac{\pi \alpha s k}{H} \gg 1$ but $\frac{\pi \alpha s k_3}{H} \ll 1$ (for the squeezed configuration), the $f_{\rm NL}$ parameters are again suppressed by large exponential factors
\begin{equation}
\begin{split}
& |f^{\rm flat}_{\rm NL}| \approx  \frac{ \lambda \dot{\phi}^2_e}{M^4} k\eta_0 e^{-2\pi \alpha s k/H} \\
& |f^{\rm sqzd}_{\rm NL}| \approx  \frac{  \lambda \dot{\phi}^2_e}{M^4}\left(\frac{k}{k_3}\right) e^{-\pi \alpha s k/H} \;\; (k \gg k_3)
\end{split}
\label{ExpSup2}
\end{equation}

\subsection{Detectability and Related Discussion}

In \cite{Dey:2012qp}, we presented a prescription for calculating the signal-to-noise of an anisotropic power spectrum of scalar fluctuations in measurements of the temperature anisotropies in the CMB. Our conservative choice of integration cutoff in that analysis left the signatures of primordial anisotropy completely beyond reach. In this work, we have scrutinized the constraints coming from back reaction and the validity of the WKB approximation. Therefore, we return to the question of detectability in the CMB power spectrum and calculate the signal-to-noise for different values of $H$ (see \cite{Dey:2012qp} and references within)
\ba\label{eq:SN2}
\left(\stn\right)^2 \equiv \langle \chi^2(C)\rangle=\int \mathrm{d}x \chi^2(C)  {\cal L}^{(iso)}= \text{Tr} \left( {C}^{(iso)} {C}^{-1} -  1 \right) +\log\left(\det {C} / \det {C}^{(iso)} \right), \;\;\;\;\;\;\;
\ab
where $C^{(iso)}$  is the isotropic component of the full covariance matrix describing the anisotropic Bianchi I metric scenario, which is given by
\begin{eqnarray}
C_{\ell \ell' m m'} \!=\! 2\delta_{m m'} \left( - i \right)^{\ell - \ell'}\!\!\!  \int k^2 dk  \Delta_\ell ( k  ) \Delta_{\ell'}^{*}( k  )  \int\limits_{-1}^1 d(\cos\theta) Y_{\ell m} \left(  \theta,\phi\!=\!0 \right) Y_{\ell' m} \left(  \theta,\phi\!=\!0 \right) P( k, \theta), \;\;\;\;\;\;\;\;
\label{corrcllmm}
\end{eqnarray}
with $P(k,\theta)$ (where $\theta\equiv \arccos(s)$) given by Eq.~(\ref{PS}).

\begin{figure}
\includegraphics[natwidth=0.50\linewidth,width=0.50\linewidth]{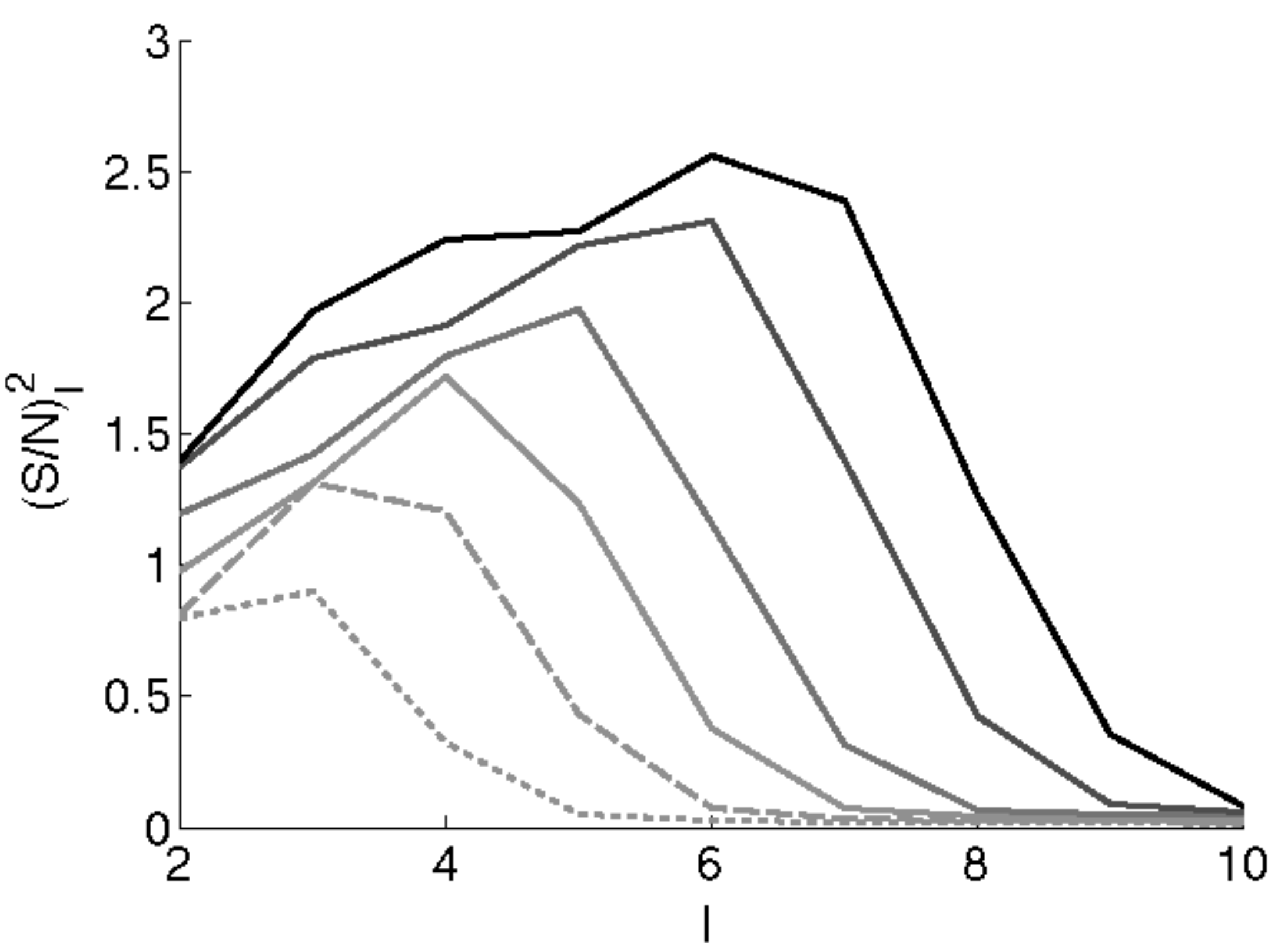}
\includegraphics[natwidth=0.47\linewidth,width=0.47\linewidth]{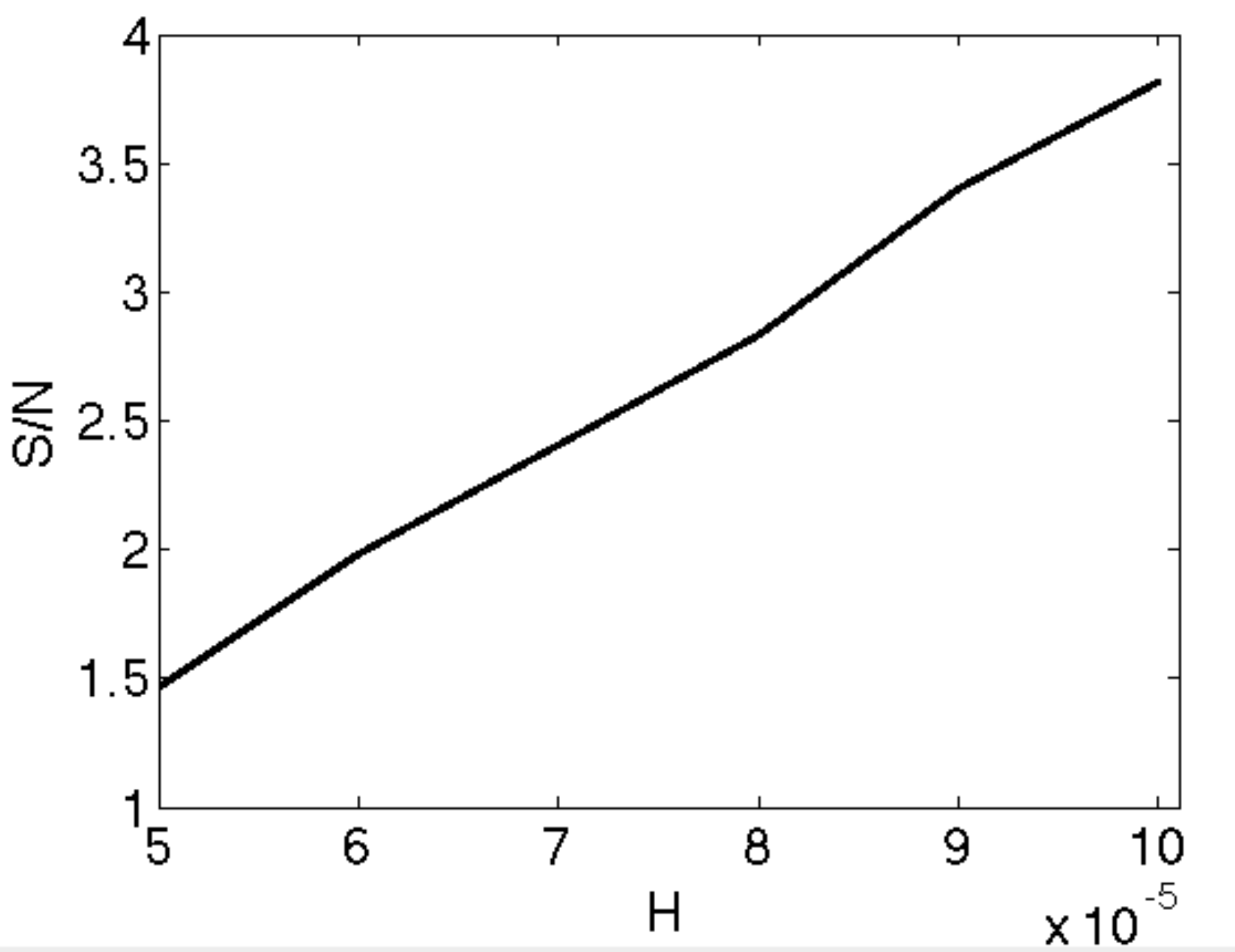}
\caption{{\it Left:} The ideal signal-to-noise per multipole $\ell$ for $H= 5\times10^{-5}-10^{-4}$, peaking at increasing values of $\ell$ (the signal is highest on the largest scales, which also have the largest cosmic variance). {\it Right:} The total signal-to-noise in $\ell\leq10$ for different values of $H$. Values of $H\sim\mathcal{O}(10^{-4})$ would be in tension with constraints on the power spectrum from the CMB. We used CMBEASY \cite{CMBEASY} to calculate the transfer functions $ \Delta_\ell ( k  )$ in Eq.~(\ref{corrcllmm}).}
\label{SN_per_l}
\end{figure}

In Fig.~\ref{SN_per_l} we plot the signal-to-noise as a function of multipole $\ell$, where the integration cutoff is determined by the conditions $(\pi \alpha s k)/H>\sqrt{k}$ and $k\gtrsim10H$ from the back reaction and WKB constraints, respectively (Eqs.~(\ref{br1}), see Section 3 for details). The anisotropic power spectrum contribution is maximized on large scales, which also have the largest cosmic variance.
While the highest possible values $H\gtrsim10^{-5}M_{\rm pl}$ are ruled out by constraints from the CMB, lower values cannot be ruled out by power spectrum measurements, even in wavenumbers within experimental reach, due to cosmic variance. In the future, with the advent of measurements of the 21-cm fluctuations from the epoch of reionization \cite{Furlanetto, Pritchard} and the dark ages \cite{LoebZald}, the number or accessible modes would increase, allowing to probe smaller values of $H$ with reasonable signal-to-noise. However, one can always tune this parameter $H$ even lower to evade those constraints as well and thus power spectrum measurements alone will not be able to rule out the full parameter space of the model considered in this paper.
  
In light of recent results from the first data release of the Planck collaboration \cite{PlanckNG}, it is useful to compare the latest experimental constraints on non-Gaussianities and those forecasted for future experiments with the enhanced signals predicted here for single-field models of inflation in the presence of primordial anisotropy. 
For the squeezed triangle configuration, the current result from Planck is $f_{\rm NL} = 2.7\pm5.8$ (at $95\%$ C.L.) \cite{PlanckNG} and with future CMB experiments this measurement error would not go down significantly. CMB experiments have access to considerably fewer modes than large scale structure surveys which in principle could be used to probe a substantial volume of redshift space within the sphere defined by the last scattering surface. The experimental prospects of using 21-cm fluctuations to constrain non-gaussianities are under debate.  Some predictions place the ultimate bound from the 21-cm fluctuations in the epoch of reionization on the order of $\Delta f_{\rm NL}\gtrsim0.2$ (see e.g.\ \cite{Joudaki, Mao}) with next generation experiments such as SKA \cite{SKA} and Omniscope \cite{Omniscope}, while more conservative estimates, incorporating the effects of foreground subtraction, predict $\Delta f_{\rm NL}\gtrsim1$ \cite{Lidz}. Futuristic all-sky experiments with access to the redshifts of the dark ages (most likely requiring an observatory on the far side of the moon \cite{Lunar}), could bring the bound as low as $\Delta f_{\rm NL}\lesssim0.01$ \cite{Cooray}, which would be enough to seriously test the vanilla single-field consistency relation against models with enhanced non-gaussianity such as described in this work. 

As mentioned in the previous section and shown below, the enhancement in $f_{\rm NL}$ as a result of the anisotropic bispectrum contribution in the scenarios we consider depends on the triangle configuration, the limits on the planarity parameter $s$, and the ratio of wavenumbers. For the squeezed shape, when $\pi \alpha s k_3/H \ll 1$ and $\pi \alpha s k/H \gtrsim 1$, exponential suppression compensates for a possible enhancement in the signal (the corresponding limit for the flattened case is when  $\pi \alpha s k/H \gg 1$). The amplitude $| f^{\rm sqzd}_{\rm NL}|$ is then $\mathcal{O}(10^{-1})$ at best, which lies below the experimental bounds from the CMB, but may be within reach of 21-cm experiments. 
In the opposite limit, contribution from the poles in the denominators of Eqs.~(\ref{squeezed1}) and (\ref{squeezed2}) may enhance the signal in particular wavenumber configurations. With standard estimators calculating the overall non-gaussian signal, these {\it local} peaks may be washed-out, but dedicated $k$-dependent estimators \cite{Hazra:2013nca} may be able to pick up this enhancement and our results motivate their consideration.

For the theory with a dimension 8 operator, Eq.~(\ref{Actionhd}), characterized by the dimensionless parameter $\lambda$, we compute the $f_{\rm NL}$ parameter at $O(\lambda)$ and note that enhancements appear again in the flattened triangle and squeezed triangle configurations.
In the limit where all wavenumbers obey $\pi \alpha s k/H \ll 1$, we predict an inverted hierarchy of amplitudes, ${|f^{\rm flat}_{\rm NL}|}>{|f^{\rm sqzd}_{\rm NL}|}$, which is particularly interesting in light of the fact that current experimental constraints on the amplitude of non-gaussianities in other triangle shapes, such as equilateral of flattened configurations, are an order-of-magnitude weaker \cite{PlanckNG}. With large scale structure data, the bounds on these shapes are expected to significantly improve in the future.

Finally, in \cite{Komatsu, Shandera}, additional observables have been suggested to detect an enhancement to the bispectrum in the squeezed limit that shows up when considering a non-BD initial state \cite{Agullo:2010ws}. In this scenario, with a simple scale-independent ansatz regarding the initial state, $f_{\rm NL}$ in the squeezed limit is enhanced by a factor $\frac{k_1}{k_3}$, where $k_1$ and $k_3$ are the small and large scales, respectively \cite{Komatsu}. 
This enhancement factor leads to two interesting signatures. The first is a distinct scale-dependence for the large scale structure halo bias (relating the power spectrum of dark matter halos with the underlying dark matter power spectrum). The second is an anisotropy in the $\mu$-type spectral distortion of the CMB that is correlated with the CMB temperature, as first demonstrated in \cite{Pajer}.

In our scenario of an anisotropic Bianchi I geometry, we calculate the scalar perturbations incorporating the full effect of the pre-inflationary dynamics, and we find that the scaling of the squeezed-limit bispectrum in the planar modes scenario contains the same $\frac{k_1}{k_3}$ enhancement factor that shows up in the non-BD initial state scenario, in addition to an oscillatory $k$-dependent term and in certain limits an exponential suppression as well.
Therefore, we also predict a significant deviation in the scale dependence of the halo bias compared to the standard $1/k^2$ resulting from the local-form bispectrum in a BD initial state (see e.g. \cite{Jeong}). The geometric oscillatory term, however, would lead to corrections to the $1/k^3$-dependence predicted in \cite{Komatsu, Shandera} for a generic non-BD ansatz. These deviations may allow to discern between different pre-inflationary scenarios, an analysis which we leave for future work.

As for the $\mu$-distortion, the reason this is a promising signature of this enhancement is that the relevant damping scale for spectral distortions in the CMB is much higher than the silk damping scale \cite{Pajer}, which allows to probe the $\frac{k_1}{k_3}$ enhancement at much larger values \cite{Komatsu}. However, this would be exploiting the limit in which $\frac{\pi \alpha s k_1}{H} \gg 1$ while $\frac{\pi \alpha s k_{3}}{H}  \ll 1$, where in our scenario, a strong exponential suppression (see Eqs.~(\ref{ExpSup1}),(\ref{ExpSup2})) overwhelms the $\frac{k_1}{k_3}$ enhancement and renders the non-gaussianity indetectable. We emphasize that a similar suppression may also arise for non-BD initial states with explicit $k$-dependent excitations, and should be taken into account in these models as well.

\section*{Acknowledgements}
SP would like to thank  the Aspen Center for Physics and the Kavli Institute for Theoretical Physics for hospitality while part of this work was done.
This material is based upon work supported by the National Science Foundation under Grant Numbers PHY-1316033 and PHY-0969020.

\appendix
\section{Standard Action: Squeezed Triangle Configuration }
In this section, we compute the $f_{\rm NL}$ parameter in the squeezed limit for generic values of the planarity parameters $s$ and $s_3$. As noted previously, the appropriate "right" amplitude is given by
\begin{equation}
A^R({\bf k},{\bf k},{\bf k_3}) \approx -\frac{k^4}{\dot{\phi}^2}\frac{H^3}{8k^6 k^3_3} \sum_{{\xi_i}=\pm1}(\prod^3_{i=1} \tilde{F_{\xi_i}}(k_i))\frac{1}{\sum_i\xi_ik_i}\left(1-e^{-i \eta_0\sum_i\xi_ik_i}\right) 
\end{equation}
where the sum is over the set   $\xi_i : (1,-1,-1),(-1,1,1),(1,-1,1),(-1,1,-1)$.\\
Using the definition of $f^{\rm sqzd}_{\rm NL}$: $f^{\rm sqzd}_{\rm NL}=\frac{A^R({\bf k},{\bf k},{\bf k_3})+c.c.}{P(k)P(k_3)} $ we have 
\begin{equation}
\begin{split}
f^{\rm sqzd}_{\rm NL}=\left(\frac{\dot{\phi}^2_e}{H^2}\right) \left(\frac{k}{k_3}\right) \frac{A(k_3, \eta_0,s_3) + e^{\pi \alpha s k/H} \, B(k,k_3,\eta_0,s_3)+ e^{-\pi \alpha s k/H}C(k,k_3,\eta_0,s_3)}{\sinh{\pi \alpha s k/H} (\cosh{\pi \alpha s k/H}-\sin{2bk/H})(\cosh{\pi \alpha s_3 k_3/H}-\sin{2bk_3/H})}
\end{split}
\end{equation}
where the functions $A(k_3, \eta_0, s_3)$ , $B(k,k_3,\eta_0,s_3)$ and $C(k,k_3,\eta_0,s_3)$ are as follows,

\begin{equation}
\begin{split}
A(k_3, \eta_0, s_3)&= -2(e^{\pi \alpha s_3 k_3/H}+ie^{i2bk_3/H})\left[(1-e^{-i\eta_0 k_3}) +i e^{-\pi \alpha s_3 k_3/H}e^{-i2bk_3/H}(1-e^{i\eta_0 k_3})\right] +c.c.\\
&=-8 \sinh{\left(\pi \alpha s_3 k_3/H\right)} \left(1-\cos{\eta_0 k_3}\right)
\end{split}
\end{equation}
\begin{equation}
\begin{split}
B(k,k_3,\eta_0,s_3)&=i(e^{\pi \alpha s_3 k_3/H}+ie^{i2bk_3/H})\left[(1-e^{-i\eta_0 k_3}) +i e^{-\pi \alpha s_3 k_3/H}e^{-i2bk_3/H}(1-e^{i\eta_0 k_3})\right] e^{-i2bk/H}+c.c.\\
&=4\sinh{\pi \alpha s_3 k_3/H}\sin{2bk/H} - 2 e^{\pi \alpha s_3 k_3/H} \sin{(2bk/H + \eta_0 k_3)} \\
&+2e^{-\pi \alpha s_3 k_3/H} \sin{(2bk/H - \eta_0 k_3)}+4\cos{2bk_3/H}\cos{2bk} -4 \cos{(\eta_0k_3-2bk_3/H)}\cos{2bk/H}  
\end{split}
\end{equation}
\begin{equation}
\begin{split}
C(k,k_3,\eta_0,s_3)&=-i(e^{\pi \alpha s_3 k_3/H}+ie^{i2bk_3/H})\left[(1-e^{-i\eta_0 k_3}) +i e^{-\pi \alpha s_3 k_3/H}e^{-i2bk_3/H}(1-e^{i\eta_0 k_3})\right] e^{i2bk/H}+c.c.\\
& = -B(-k,k_3,\eta_0,s_3)
\end{split}
\end{equation}
Therefore, the formula for $f^{\rm sqzd}_{\rm NL}$ may be re-written as
\begin{equation}
\begin{split}
&f^{\rm sqzd}_{\rm NL}=\left(\frac{\dot{\phi}^2_e}{H^2}\right) \left(\frac{k}{k_3}\right) \frac{F(k,k_3, \eta_0,s_3)}{\sinh{\pi \alpha s k/H} (\cosh{\pi \alpha s k/H}-\sin{2bk/H})(\cosh{\pi \alpha s_3 k_3/H}-\sin{2bk_3/H})}\\
&F(k,k_3, \eta_0,s_3)=4\Big[2\sinh{\pi \alpha s_3 k_3/H}\sin{2bk/H}\cos{\pi \alpha s k/H}\\
&- \sin{(2bk/H+\eta_0k_3)}\sinh{(\pi \alpha s k/H+\pi \alpha s_3 k_3/H)} + \sin{(2bk/H-\eta_0k_3)}\sinh{(\pi \alpha s k/H-\pi \alpha s_3 k_3/H)} \\
&-2\cos{2bk_3/H}\cos{2bk/H}\sinh{\pi \alpha s k/H}+2\cos{(2bk_3/H- \eta_0 k_3)}\cos{2bk/H}\sinh{\pi \alpha s k/H}\Big]
\end{split}
\end{equation}

\section{Higher Derivative Operator: Flattened Triangle Configuration}

For the flattened triangle configuration, we choose $k_3 \approx k_2 \approx k_1/2 \approx k$, which sets $k_1=k_2+k_3$, such that the integrands appearing in the "right" amplitudes $A^{(1)}_R$ and $A^{(2)}_R$ in equation (\ref{rightamp1}) and equation (\ref{rightamp2}) respectively have contributions only from the following configurations : $\{+,-,-\}$ and $\{-,+,+\}$. 
Therefore, the two contributions for the right amplitude in this case are given as follows,
\begin{equation}
\begin{split}
&A^{(1)}_R({\bf k_1},{\bf k_2},{\bf k_3})+c.c.=-i\int^{0}_{\eta_0}d\eta  \frac{\lambda \dot{\phi}^4}{\dot{\rho}^4 M^4} (\frac{\dot{\rho}^{12}}{\dot{\phi}^{6}})\frac{\eta^2}{8k_1k_2 k_3} \Big[\tilde{F}_{+}({\bf k_1})\tilde{F}_{-}({\bf k_2})\tilde{F}_{-}({\bf k_3})e^{-i(k_1-k_2-k_3)\eta}\\
&+\tilde{F}_{-}({\bf k_1})\tilde{F}_{+}({\bf k_2})\tilde{F}_{+}({\bf k_3})e^{-i(-k_1+k_2+k_3)\eta}\Big]\times (3!) +c.c.\\
=& \frac{\lambda \dot{\phi}^4}{\dot{\rho}^4 M^4} (\frac{\dot{\rho}^{12}}{\dot{\phi}^{6}}) \frac{\eta_0^3}{16 k^3}\frac{\Big[ \sinh{\frac{2\pi \alpha s k}{H}} (\cos{4bk/H}-\sin{4bk/H})+2\sinh{\frac{\pi \alpha s k}{H}} (\cos{2bk/H}+\sin{2bk/H}) \Big]}{\sinh{\frac{2\pi \alpha s k}{H}} \sinh^2{\frac{\pi \alpha s k}{H}}}\\
=&\frac{\lambda \dot{\phi}^4}{\dot{\rho}^4 M^4} (\frac{\dot{\rho}^{12}}{\dot{\phi}^{6}}) \frac{\eta_0^3}{8 k^3}\frac{\Big[ \cosh{\frac{\pi \alpha s k}{H}} (\cos{4bk/H}-\sin{4bk/H})+ (\cos{2bk/H}+\sin{2bk/H}) \Big]}{\sinh{\frac{2\pi \alpha s k}{H}} \sinh{\frac{\pi \alpha s k}{H}}} \label{A1-flat}
\end{split}
\end{equation}

\begin{equation}
\begin{split}
A^{(2)}_R({\bf k_1},{\bf k_2},{\bf k_3})=&-i\int^{0}_{\eta_0}d\eta  \frac{\lambda \dot{\phi}^4}{\dot{\rho}^4 M^4} (\frac{\dot{\rho}^{6}}{\dot{\phi}^{6}})\prod^3_{i=1}\frac{\dot{\rho}^{2}}{2k^3_i} [\tilde{F}_{+}({\bf k_1})\tilde{F}_{-}({\bf k_2})\tilde{F}_{-}({\bf k_3})e^{-i(k_1-k_2-k_3)\eta}\mathcal{A}({\bf k_1},{\bf k_2},{\bf k_3},\eta)\\
&+\tilde{F}_{-}({\bf k_1})\tilde{F}_{+}({\bf k_2})\tilde{F}_{+}({\bf k_3})e^{-i(-k_1+k_2+k_3)\eta}\mathcal{B}({\bf k_1},{\bf k_2},{\bf k_3},\eta)]\times (2!) 
\end{split}
\end{equation}
where the two functions $A({\bf k_1},{\bf k_2},{\bf k_3},\eta)$ and $B({\bf k_1},{\bf k_2},{\bf k_3},\eta)$ are given as,
\begin{equation}
\begin{split}
\mathcal{A}({\bf k_1},{\bf k_2},{\bf k_3},\eta)&=(\vec{k_1}.\vec{k_2})k^2_3(1+ik_1\eta)(1-ik_2\eta) +(\vec{k_1}.\vec{k_3})k^2_2(1+ik_1\eta)(1-ik_3\eta) \\
&+(\vec{k_2}.\vec{k_3})k^2_1(1-ik_2\eta)(1-ik_3\eta) \\
&= -12k^4\left[ik\eta+k^2\eta^2 \right]
\end{split}
\end{equation}
\begin{equation}
\begin{split}
\mathcal{B}({\bf k_1},{\bf k_2},{\bf k_3},\eta)=&(\vec{k_1}.\vec{k_2})k^2_3(1-ik_1\eta)(1+ik_2\eta) +(\vec{k_1}.\vec{k_3})k^2_2(1-ik_1\eta)(1+ik_3\eta) \\
&+(\vec{k_2}.\vec{k_3})k^2_1(1+ik_2\eta)(1+ik_3\eta) \\
=&-12k^4 \left[-ik\eta+k^2\eta^2 \right]
\end{split}
\end{equation}

Therefore, we have 
\begin{equation}
\begin{split}
&A^{(2)}_R({\bf k_1},{\bf k_2},{\bf k_3})+c.c.\\
=&\frac{\lambda \dot{\phi}^4}{\dot{\rho}^4 M^4} (\frac{\dot{\rho}^{6}}{\dot{\phi}^{6}})\frac{3\dot{\rho}^{6}}{8 k^5} \frac{\Big[ \cosh{\frac{\pi \alpha s k}{H}} (\cos{\frac{4bk}{H}}-\sin{\frac{4bk}{H}})+ (\cos{\frac{2bk}{H}}+\sin{\frac{2bk}{H}}) \Big](k\eta^2_0-\frac{2k^2 \eta^3_0}{3})}{2\sinh{\frac{2\pi \alpha s k}{H}} \sinh{\frac{\pi \alpha s k}{H}}} \label{A2-flat}
\end{split}
\end{equation}
From equation (\ref{A1-flat}) and equation (\ref{A2-flat}), we have
\begin{equation}
\begin{split}
&A^R({\bf k},{\bf k},{\bf k}) + c.c.\\
&=A^{(1)}_R({\bf k_1},{\bf k_2},{\bf k_3})+A^{(2)}_R({\bf k_1},{\bf k_2},{\bf k_3}) +c.c.\\
&=\frac{\lambda \dot{\phi}^4}{\dot{\rho}^4 M^4} (\frac{\dot{\rho}^{6}}{\dot{\phi}^{6}})\frac{3\dot{\rho}^{6}}{8 k^6} \frac{\Big(\cosh{\frac{\pi \alpha s k}{H}} (\cos{\frac{4bk}{H}}-\sin{\frac{4bk}{H}}) \Big)(k^2\eta^2_0)}{2\sinh{\frac{2\pi \alpha s k}{H}} \sinh{\frac{\pi \alpha s k}{H}}}
\end{split}
\end{equation}
Using the definition of $f^{\rm flat}_{\rm NL}: f^{\rm flat}_{\rm NL}=(A^R({\bf k},{\bf k},{\bf k}) + c.c.)/P({\bf k})^2$ and taking off a factor of $k\eta_0$ to account for the 2D projection of the naive $f_{\rm NL}$, we have
\begin{equation}
f^{\rm flat}_{\rm NL}=- \frac{3 \lambda \dot{\phi}^2_e}{16 M^4} k\eta_0 \left(\frac{\cos{(4bk/H)} -\sin{(4bk/H)}}{\left(\cosh{(\pi \alpha s k/H)}-\sin{(2bk/H)}\right)^2}\right)
\end{equation}

\section{Higher Derivative Operator: Squeezed Triangle Configuration}
For the squeezed triangle configuration, we choose $k_3 \ll k_1 \approx k_2 \sim k$ and therefore the configurations that contribute to the integrands  of the "right" amplitudes $A^{(1)}_R$ and $A^{(2)}_R$ at the leading order are $\{+,-,-\},\{-,+,+\},\{+,-,+\},\{-,+,-\}$.
Explicitly, the contributions to the right amplitudes are
\begin{equation}
\begin{split}
&A^{(1)}_R({\bf k_1},{\bf k_2},{\bf k_3})+c.c.=-i\int^{0}_{\eta_0}d\eta  \frac{\lambda \dot{\phi}^4}{\dot{\rho}^4 M^4} (\frac{\dot{\rho}^{12}}{\dot{\phi}^{6}})\frac{\eta^2}{8k_1k_2 k_3} [\tilde{F}_{+}({\bf k_1})\tilde{F}_{-}({\bf k_2})\tilde{F}_{-}({\bf k_3})e^{-i(k_1-k_2-k_3)\eta}\\
&+\tilde{F}_{-}({\bf k_1})\tilde{F}_{+}({\bf k_2})\tilde{F}_{+}({\bf k_3})e^{-i(-k_1+k_2+k_3)\eta}+\tilde{F}_{+}({\bf k_1})\tilde{F}_{-}({\bf k_2})\tilde{F}_{+}({\bf k_3})e^{-i(k_1-k_2+k_3)\eta}\\
&+\tilde{F}_{-}({\bf k_1})\tilde{F}_{+}({\bf k_2})\tilde{F}_{-}({\bf k_3})e^{-i(-k_1+k_2-k_3)\eta}]\times (3!) +c.c.\\
&=\frac{\lambda \dot{\phi}^4}{\dot{\rho}^4 M^4} (\frac{\dot{\rho}^{12}}{\dot{\phi}^{6}}) \frac{3}{2k^2k^4_3}\left( \frac{\sin{\frac{k_3\eta_0}{2}}\left(\sinh{\frac{\pi \alpha s k}{H}}U(k,k_3,\eta_0)-\sinh{\frac{\pi \alpha s k_3}{H}}V(k,k_3,\eta_0)\right)}{\sinh^2{\frac{\pi \alpha s k}{H}}\sinh{\frac{\pi \alpha s k_3}{H}}} \right)
\end{split}
\end{equation}
where the functions $U$ and $V$ are given as
\begin{equation}
\begin{split}
&U(k,k_3,\eta_0)= \cos{(\frac{2bk}{H}+\frac{k_3\eta_0}{2})} -\sin{(\frac{2bk}{H}+\frac{k_3\eta_0}{2})}+\cos{(\frac{2bk}{H}-\frac{k_3\eta_0}{2})}+\sin{(\frac{2bk}{H}-\frac{k_3\eta_0}{2})}\\
&V(k,k_3,\eta_0)=2\sin{\frac{k_3\eta_0}{2}} 
\end{split}
\end{equation}

\begin{equation}
\begin{split}
&A^{(2)}_R({\bf k_1},{\bf k_2},{\bf k_3})=-i\int^{0}_{\eta_0}d\eta  \frac{\lambda \dot{\phi}^4}{\dot{\rho}^4 M^4} (\frac{\dot{\rho}^{6}}{\dot{\phi}^{6}})\prod^3_{i=1}\frac{\dot{\rho}^{2}}{2k^3_i} [\tilde{F}_{+}({\bf k_1})\tilde{F}_{-}({\bf k_2})\tilde{F}_{-}({\bf k_3})e^{-i(k_1-k_2-k_3)\eta}\mathcal{A}_1({\bf k_1},{\bf k_2},{\bf k_3},\eta)\\
&+\tilde{F}_{-}({\bf k_1})\tilde{F}_{+}({\bf k_2})\tilde{F}_{+}({\bf k_3})e^{-i(-k_1+k_2+k_3)\eta}\mathcal{A}_2({\bf k_1},{\bf k_2},{\bf k_3},\eta)+\tilde{F}_{+}({\bf k_1})\tilde{F}_{-}({\bf k_2})\tilde{F}_{+}({\bf k_3})e^{-i(k_1-k_2+k_3)\eta}\mathcal{A}_3({\bf k_1},{\bf k_2},{\bf k_3},\eta)\\
&+\tilde{F}_{-}({\bf k_1})\tilde{F}_{+}({\bf k_2})\tilde{F}_{-}({\bf k_3})e^{-i(-k_1+k_2-k_3)\eta}\mathcal{A}_4({\bf k_1},{\bf k_2},{\bf k_3},\eta)]\times (2!) 
\end{split}
\end{equation}
As before, the functions $\mathcal{A}_i({\bf k_1},{\bf k_2},{\bf k_3},\eta)$ are given as,
\begin{equation}
\begin{split}
\mathcal{A}_1({\bf k_1},{\bf k_2},{\bf k_3},\eta)=&(\vec{k_1}.\vec{k_2})k^2_3(1+ik_1\eta)(1-ik_2\eta) +(\vec{k_1}.\vec{k_3})k^2_2(1+ik_1\eta)(1-ik_3\eta) \\
&+(\vec{k_2}.\vec{k_3})k^2_1(1-ik_2\eta)(1-ik_3\eta) \\
\approx & -k^2k_3^2 (2 + k^2\eta^2 -i k_3\eta)\\
\mathcal{A}_2({\bf k_1},{\bf k_2},{\bf k_3},\eta)=&(\vec{k_1}.\vec{k_2})k^2_3(1-ik_1\eta)(1+ik_2\eta) +(\vec{k_1}.\vec{k_3})k^2_2(1-ik_1\eta)(1+ik_3\eta) \\
&+(\vec{k_2}.\vec{k_3})k^2_1(1+ik_2\eta)(1+ik_3\eta) \\
\approx & -k^2k_3^2 (2 + k^2\eta^2 +i k_3\eta)\\
\mathcal{A}_3({\bf k_1},{\bf k_2},{\bf k_3},\eta)=&(\vec{k_1}.\vec{k_2})k^2_3(1+ik_1\eta)(1-ik_2\eta) +(\vec{k_1}.\vec{k_3})k^2_2(1+ik_1\eta)(1+ik_3\eta) \\
&+(\vec{k_2}.\vec{k_3})k^2_1(1-ik_2\eta)(1+ik_3\eta) \\
\approx &  -k^2k_3^2 (2 + k^2\eta^2 +i k_3\eta)\\
\mathcal{A}_4({\bf k_1},{\bf k_2},{\bf k_3},\eta)=&(\vec{k_1}.\vec{k_2})k^2_3(1-ik_1\eta)(1+ik_2\eta) +(\vec{k_1}.\vec{k_3})k^2_2(1-ik_1\eta)(1-ik_3\eta) \\
&+(\vec{k_2}.\vec{k_3})k^2_1(1+ik_2\eta)(1-ik_3\eta) \\
\approx &  -k^2k_3^2 (2 + k^2\eta^2 -i k_3\eta)
\end{split}
\end{equation}
The leading order contribution to $A^{(2)}_R({\bf k_1},{\bf k_2},{\bf k_3})$ for $k \gg k_3$ can be shown to be
\begin{equation}
\begin{split}
A^{(2)}_R({\bf k_1},{\bf k_2},{\bf k_3}) \approx -\frac{1}{3}A^{(1)}_R({\bf k_1},{\bf k_2},{\bf k_3})
\end{split}
\end{equation}

Now, from the definition$f^{\rm sqzd}_{\rm NL}$: $f^{\rm sqzd}_{\rm NL}=\frac{A^R({\bf k},{\bf k},{\bf k_3})+c.c.}{P(k)P(k_3)} =\frac{A^{(1)}_R({\bf k},{\bf k},{\bf k_3})+A^{(2)}_R({\bf k},{\bf k},{\bf k_3})+c.c.}{P(k)P(k_3)} $, we have
\begin{equation}
\begin{split}
\boxed{f^{\rm sqzd}_{\rm NL}= \frac{ 4 \lambda \dot{\phi}^2_e}{M^4}\left(\frac{k}{k_3}\right) \left(\frac{\sin{\frac{k_3\eta_0}{2}}\left(\sinh{\frac{\pi \alpha s k}{H}}U(k,k_3,\eta_0)-\sinh{\frac{\pi \alpha s k_3}{H}}V(k,k_3,\eta_0)\right)}{\sinh{\pi \alpha s k/H}\left(\cosh{(\pi \alpha s k/H)}-\sin{(2bk/H)}\right)\left(\cosh{(\pi \alpha s k_3/H)}-\sin{(2bk_3/H)}\right)}\right)}
\end{split}
\end{equation}

\end{document}